\input harvmac

%\DineZA
\lref\DineZA{
  M.~Dine, W.~Fischler and M.~Srednicki,
  ``Supersymmetric Technicolor,''
  Nucl.\ Phys.\ B {\bf 189}, 575 (1981).
  %%CITATION = NUPHA,B189,575;%%
}

%\DimopoulosAU
\lref\DimopoulosAU{
  S.~Dimopoulos and S.~Raby,
  ``Supercolor,''
  Nucl.\ Phys.\ B {\bf 192}, 353 (1981).
  %%CITATION = NUPHA,B192,353;%%
}

%\DineGU
\lref\DineGU{
  M.~Dine and W.~Fischler,
  ``A Phenomenological Model Of Particle Physics Based On Supersymmetry,''
  Phys.\ Lett.\ B {\bf 110}, 227 (1982).
  %%CITATION = PHLTA,B110,227;%%
}

%\NappiHM
\lref\NappiHM{
  C.~R.~Nappi and B.~A.~Ovrut,
  ``Supersymmetric Extension Of The SU(3) X SU(2) X U(1) Model,''
  Phys.\ Lett.\ B {\bf 113}, 175 (1982).
  %%CITATION = PHLTA,B113,175;%%
}

%\AlvarezGaumeWY
\lref\AlvarezGaumeWY{
  L.~Alvarez-Gaume, M.~Claudson and M.~B.~Wise,
  ``Low-Energy Supersymmetry,''
  Nucl.\ Phys.\ B {\bf 207}, 96 (1982).
  %%CITATION = NUPHA,B207,96;%%
}

%\DimopoulosGM
\lref\DimopoulosGM{
  S.~Dimopoulos and S.~Raby,
  ``Geometric Hierarchy,''
  Nucl.\ Phys.\ B {\bf 219}, 479 (1983).
  %%CITATION = NUPHA,B219,479;%%
}

%\DineYW
\lref\DineYW{
  M.~Dine and A.~E.~Nelson,
  ``Dynamical supersymmetry breaking at low-energies,''
  Phys.\ Rev.\ D {\bf 48}, 1277 (1993)
  [arXiv:hep-ph/9303230].
  %%CITATION = HEP-PH 9303230;%%
}

%\BanksMA
\lref\BanksMA{
  T.~Banks,
  ``Remodeling the pentagon after the events of 2/23/06,''
  arXiv:hep-ph/0606313.
  %%CITATION = HEP-PH 0606313;%%
}

%\DineVC
\lref\DineVC{
  M.~Dine, A.~E.~Nelson and Y.~Shirman,
  ``Low-Energy Dynamical Supersymmetry Breaking Simplified,''
  Phys.\ Rev.\ D {\bf 51}, 1362 (1995)
  [arXiv:hep-ph/9408384].
  %%CITATION = HEP-PH 9408384;%%
}

%\DineAG
\lref\DineAG{
  M.~Dine, A.~E.~Nelson, Y.~Nir and Y.~Shirman,
  ``New tools for low-energy dynamical supersymmetry breaking,''
  Phys.\ Rev.\ D {\bf 53}, 2658 (1996)
  [arXiv:hep-ph/9507378].
  %%CITATION = HEP-PH 9507378;%%
}

%\GiudiceBP
\lref\GiudiceBP{
  G.~F.~Giudice and R.~Rattazzi,
  ``Theories with gauge-mediated supersymmetry breaking,''
  Phys.\ Rept.\  {\bf 322}, 419 (1999)
  [arXiv:hep-ph/9801271].
  %%CITATION = HEP-PH 9801271;%%
}

%\IntriligatorDD
\lref\IntriligatorDD{
  K.~Intriligator, N.~Seiberg and D.~Shih,
  ``Dynamical SUSY breaking in meta-stable vacua,''
  JHEP {\bf 0604}, 021 (2006)
  [arXiv:hep-th/0602239].
  %%CITATION = HEP-TH 0602239;%%
}

%\FrancoES
\lref\FrancoES{
  S.~Franco and A.~M.~Uranga,
  ``Dynamical SUSY breaking at meta-stable minima from D-branes at obstructed
  geometries,''
  JHEP {\bf 0606}, 031 (2006)
  [arXiv:hep-th/0604136].
  %%CITATION = HEP-TH 0604136;%%
}

%\KitanoWM
\lref\KitanoWM{
  R.~Kitano,
  ``Dynamical GUT breaking and mu-term driven supersymmetry breaking,''
  arXiv:hep-ph/0606129.
  %%CITATION = HEP-PH 0606129;%%
}

%\OoguriPJ
\lref\OoguriPJ{
  H.~Ooguri and Y.~Ookouchi,
  ``Landscape of supersymmetry breaking vacua in geometrically realized gauge
  theories,''
  Nucl.\ Phys.\ B {\bf 755}, 239 (2006)
  [arXiv:hep-th/0606061].
  %%CITATION = HEP-TH 0606061;%%
}

%\KitanoWZ
\lref\KitanoWZ{
  R.~Kitano,
  ``Gravitational gauge mediation,''
  Phys.\ Lett.\ B {\bf 641}, 203 (2006)
  [arXiv:hep-ph/0607090].
  %%CITATION = HEP-PH 0607090;%%
}

%\AmaritiVK
\lref\AmaritiVK{
  A.~Amariti, L.~Girardello and A.~Mariotti,
  ``Non-supersymmetric meta-stable vacua in SU(N) SQCD with adjoint matter,''
  arXiv:hep-th/0608063.
  %%CITATION = HEP-TH 0608063;%%
}

%\DineGM
\lref\DineGM{
  M.~Dine, J.~L.~Feng and E.~Silverstein,
  ``Retrofitting O'Raifeartaigh models with dynamical scales,''
  Phys.\ Rev.\ D {\bf 74}, 095012 (2006)
  [arXiv:hep-th/0608159].
  %%CITATION = HEP-TH 0608159;%%
}

%\SeibergBZ
\lref\SeibergBZ{
  N.~Seiberg,
  ``Exact Results On The Space Of Vacua Of Four-Dimensional Susy Gauge
  Theories,''
  Phys.\ Rev.\ D {\bf 49}, 6857 (1994)
  [arXiv:hep-th/9402044].
  %%CITATION = HEP-TH 9402044;%%
}

%\WittenNF
\lref\WittenNF{
  E.~Witten,
  ``Dynamical Breaking Of Supersymmetry,''
  Nucl.\ Phys.\ B {\bf 188}, 513 (1981).
  %%CITATION = NUPHA,B188,513;%%
}

%\DineXT
\lref\DineXT{
  M.~Dine and J.~Mason,
  ``Gauge mediation in metastable vacua,''
  arXiv:hep-ph/0611312.
  %%CITATION = HEP-PH 0611312;%%
}

%\KitanoXG
\lref\KitanoXG{
  R.~Kitano, H.~Ooguri and Y.~Ookouchi,
  ``Direct mediation of meta-stable supersymmetry breaking,''
  arXiv:hep-ph/0612139.
  %%CITATION = HEP-PH 0612139;%%
}

%\MurayamaYF
\lref\MurayamaYF{
  H.~Murayama and Y.~Nomura,
  ``Gauge Mediation Simplified,''
  arXiv:hep-ph/0612186.
  %%CITATION = HEP-PH 0612186;%%
}
%\CsakiWI
\lref\CsakiWI{
  C.~Csaki, Y.~Shirman and J.~Terning,
  ``A Simple Model of Low-scale Direct Gauge Mediation,''
  arXiv:hep-ph/0612241.
  %%CITATION = HEP-PH 0612241;%%
}

%\MurayamaPB
\lref\MurayamaPB{
  H.~Murayama,
  ``A model of direct gauge mediation,''
  Phys.\ Rev.\ Lett.\  {\bf 79}, 18 (1997)
  [arXiv:hep-ph/9705271].
  %%CITATION = HEP-PH 9705271;%%
}

%\DimopoulosWW
\lref\DimopoulosWW{
  S.~Dimopoulos, G.~R.~Dvali, R.~Rattazzi and G.~F.~Giudice,
  ``Dynamical soft terms with unbroken supersymmetry,''
  Nucl.\ Phys.\ B {\bf 510}, 12 (1998)
  [arXiv:hep-ph/9705307].
  %%CITATION = HEP-PH 9705307;%%
}

%\LutyNY
\lref\LutyNY{
  M.~A.~Luty,
  ``Simple gauge-mediated models with local minima,''
  Phys.\ Lett.\ B {\bf 414}, 71 (1997)
  [arXiv:hep-ph/9706554].
  %%CITATION = HEP-PH 9706554;%%
}

%\DimopoulosJE
\lref\DimopoulosJE{
  S.~Dimopoulos, G.~R.~Dvali and R.~Rattazzi,
  ``A simple complete model of gauge-mediated SUSY-breaking and dynamical
  relaxation mechanism for solving the mu problem,''
  Phys.\ Lett.\ B {\bf 413}, 336 (1997)
  [arXiv:hep-ph/9707537].
  %%CITATION = HEP-PH 9707537;%%
}

%\AffleckXZ
\lref\AffleckXZ{
  I.~Affleck, M.~Dine and N.~Seiberg,
  ``Dynamical Supersymmetry Breaking In Four-Dimensions And Its
  Phenomenological Implications,''
  Nucl.\ Phys.\ B {\bf 256}, 557 (1985).
  %%CITATION = NUPHA,B256,557;%%
}

%\SeibergPQ
\lref\SeibergPQ{
  N.~Seiberg,
  ``Electric - magnetic duality in supersymmetric nonAbelian gauge theories,''
  Nucl.\ Phys.\ B {\bf 435}, 129 (1995)
  [arXiv:hep-th/9411149].
  %%CITATION = HEP-TH 9411149;%%
}

%\NelsonNF
\lref\NelsonNF{
  A.~E.~Nelson and N.~Seiberg,
  ``R symmetry breaking versus supersymmetry breaking,''
  Nucl.\ Phys.\ B {\bf 416}, 46 (1994)
  [arXiv:hep-ph/9309299].
  %%CITATION = HEP-PH 9309299;%%
}

\def\K3{{\bf K3}}
\def\journal#1&#2(#3){\unskip, \sl #1\ \bf #2 \rm(19#3) }
\def\andjournal#1&#2(#3){\sl #1~\bf #2 \rm (19#3) }

\def\tilde{\widetilde}

\def\frac#1#2{{#1\over#2}}

\def\inbar{\,\vrule height1.5ex width.4pt depth0pt}
\def\IC{\relax\hbox{$\inbar\kern-.3em{\rm C}$}}
\def\IR{\relax{\rm I\kern-.18em R}}
\def\IP{\relax{\rm I\kern-.18em P}}

%
%%%%%%%%%%%%%%%%%%%%%%%%%%%%%%%%%%%%
%

%
\catcode`\@=11
\def\slash#1{\mathord{\mathpalette\c@ncel{#1}}}
\overfullrule=0pt

\def\underrel#1\over#2{\mathrel{\mathop{\kern\z@#1}\limits_{#2}}}

\catcode`\@=12

%%%%%%%%%%%%%%%%%%%%%%%%%%%%%%%%%%%%%%%%%%%%%%%%%%%%%%%%%%%%%%

%

%%%%%%%%%%%%%%%%%%%%%%%%%%%%%%%%%%%%%%%%%%%%%%%%%%%%%%%%%%%%%%
% new defs:

%%%%%%%%%%%%%%%%%%%%%%%%%%%%%%%%%%%%%%%%%%%%%%%%%%%

\Title{\rightline{WIS/19/06-DEC-DPP} }
{\vbox{\centerline{Naturalized and simplified gauge mediation}
 }}
\medskip

\centerline{\it Ofer Aharony$^{1,2}$ and Nathan Seiberg$^1$}
\bigskip
\centerline{$^1$School of Natural Sciences} \centerline{Institute
for Advanced Study} \centerline{Einstein Drive, Princeton, NJ 08540,
USA}
\bigskip
\centerline{$^2$Department of Particle Physics} \centerline{Weizmann
Institute of Science} \centerline{Rehovot 76100, Israel}

\smallskip

\vglue .3cm

\bigskip
\bigskip
\noindent

Following recent developments in model building we construct a
simple, natural and controllable model of gauge-mediated
supersymmetry breaking.

\Date{December 2006}

%\draftmode

It is an important challenge to find an explicit model of dynamical
supersymmetry breaking \WittenNF\ which satisfies all known
experimental constraints.  The most promising avenue is that of
gauge-mediated supersymmetry breaking
\refs{\DineZA\DimopoulosAU\DineGU\NappiHM\AlvarezGaumeWY\DimopoulosGM
\DineYW\DineVC-\DineAG} (for a review, see e.g.\ \GiudiceBP).

Unfortunately, all these models are extremely complicated.  The
complexity originates from the sector of the theory which
dynamically breaks supersymmetry and is made worse by the
messengers.

Following \IntriligatorDD, many authors have recently found
extremely simple models of supersymmetry breaking in a metastable
state
\refs{\IntriligatorDD\FrancoES\OoguriPJ\KitanoWM\BanksMA\KitanoWZ\AmaritiVK
\DineGM\DineXT\KitanoXG\MurayamaYF-\CsakiWI} (for earlier work see
e.g.\ \refs{\DineYW\DineVC-\DineAG,\MurayamaPB\DimopoulosWW
\LutyNY-\DimopoulosJE}).  One can hope to use the large global
symmetry of these simple models to construct a model of direct
gauge mediation in which no messengers are needed \AffleckXZ, as
was recently discussed in
\refs{\IntriligatorDD,\BanksMA,\KitanoXG,\CsakiWI}.  Here,
following \MurayamaYF, we will explore a gauge mediation model
with messengers.

The main point of this paper is to make the model of \MurayamaYF\
completely natural.  That model has a number of mass scales which
were put in ``by hand."  In a completely natural model such scales
should arise dynamically.  This is easily achieved, as in
\refs{\DineGM,\DineXT}, by introducing another gauge group and
replacing every dimensionful constant by the gluino bilinear of
that group divided by an appropriate power of a high scale, which
we will take to be the Planck scale $M_p$.

Our model is extremely simple and completely natural.  It also has
many simple variants.  We do not necessarily advocate this model
as a model of Nature, but merely as an illustration that the
traditional ideas of naturalness and dynamical supersymmetry
breaking can be realized in a simple and economical model.

Our model includes three sectors:
 \item{1.} The first sector is a supersymmetric
Yang-Mills theory with some non-Abelian gauge group $G$ with no
charged fields. It is characterized by a scale $\tilde \Lambda$
where it becomes strong. The purpose of this sector is to
dynamically generate the mass parameters in the Lagrangian of
\MurayamaYF, as in \refs{\DineGM,\DineXT}.  The scale $\tilde
\Lambda$ is larger than any other scale in the problem except
$M_p$, and therefore the dynamics of this sector should be
analyzed first.
 \item{2.} The second sector breaks supersymmetry.  Here we follow
 \IntriligatorDD\ and consider the supersymmetric QCD (SQCD)
 theory with gauge group
 $SU(N_c)$ and with $N_f$ quarks $Q_i$,
 $\tilde Q^j$, with $N_c+1 \leq N_f < 3N_c/2$, characterized by a
scale $\Lambda$.  Clearly, this theory can be replaced by other
theories which similarly break supersymmetry.  We do not include
mass terms for the quarks in this sector.  These arise only through
the coupling to the first sector via high dimension operators.
 \item{3.} The third sector includes the MSSM (or some other
 supersymmetric generalization of the standard model), coupled to
 messenger fields
$f$ and $\tilde f$ (which can be, for instance, in the $\bf 5$ and
$\bf \overline 5$ of an $SU(5)$ GUT group).  We do not include
explicit mass terms for the messengers\foot{With the exception of
the $\mu$ parameter which we do not discuss, the gauge symmetries
forbid mass terms for the MSSM fields.}. Such a mass term arises
from the coupling to the first sector via high dimension
operators.

Suppressing factors of order one, the terms in the action which
are relevant for our analysis are of the form
 \eqn\superpot{\int d^2 \theta \left[ {1\over M_p} Q_i \tilde Q^i
 f \tilde f
 + \Tr (W_\alpha^2)  \left({1 \over g_{SYM}^2}+ {1 \over M_p^2} (
 Q_i \tilde Q^i + f \tilde f) \right) \right],}
where $W_{\alpha}$ is the chiral multiplet including the gauge field
of the group $G$. Here we assumed that the MSSM fields do not couple
to $Q$ and $W_\alpha$ at the leading order\foot{This can be made
natural with an appropriate symmetry. Without such a symmetry this
is automatic for one of the couplings in \superpot\ (this is the
definition of the messenger fields), but we assume that it is true
for both.}.  In addition to \superpot\ there are standard terms for
the MSSM fields, and for the SQCD sector which breaks supersymmetry,
and canonical K\"ahler terms. For simplicity we assumed that the
action preserves an $SU(N_f)$ flavor symmetry of the SQCD quarks.
This assumption is not essential, and our conclusions do not change
if arbitrary $SU(N_f)$-breaking couplings are allowed in \superpot.

We take
 \eqn\siinequ{\Lambda \ll \tilde \Lambda \ll M_p.}
The Planck scale $M_p$ may be replaced by any other high-energy
scale which we do not discuss in detail. Note that the
superpotential \superpot\ does not have a continuous R-symmetry
(but just an R-parity symmetry).

The largest scale in the model is $\tilde \Lambda$.  Here, the
gauge theory of $G$ becomes strong. Gaugino condensation in this
theory leads to $\langle \Tr(W_\alpha^2) \rangle \sim \tilde
\Lambda^3$, generating masses for the SQCD quarks $Q$ and the
messengers $f$
 \eqn\massesQf{m_Q \sim m_f \sim {\tilde \Lambda^3 \over M_p^2}.}

The next scale we encounter is the scale $\Lambda$ where the SQCD
theory becomes strongly coupled. We will take
 \eqn\mqlam{m_Q \ll \Lambda .}
Then, below the scale $\Lambda$ the weakly coupled degrees of
freedom are those of the dual ``magnetic" theory \SeibergPQ.  In
terms of these degrees of freedom we have
 \eqn\superpote{W = \Phi_i^j q^i \tilde q_j + m_Q \Lambda\Tr(\Phi)
 +  ( m_f + {\Lambda \over M_p} \Tr
 (\Phi)) f \tilde f, }
where $q$ and $\tilde q$ are the dual quarks (or the baryons if
$N_f = N_c+1$ \SeibergBZ) and $\Phi_i^j \equiv Q_i \tilde Q^j /
\Lambda$. As discussed in \IntriligatorDD, when $m_Q \ll \Lambda$
this theory has a metastable SUSY-breaking vacuum, in which the
$F$-term of $\Phi$ is of order $F_{\Phi}\sim m_Q \Lambda$.

The low energy spectrum in this vacuum has a number of massless
fields including Goldstone bosons of some global symmetries and some
of the fermionic components of $\Phi$ \IntriligatorDD.  These can be
made massive by gauging the global symmetry, by including
$SU(N_f)$-violating couplings in \superpot, or by including other
higher dimension operators \MurayamaYF.

In this supersymmetry breaking vacuum, the effective superpotential
for the messenger fields well below the scale $\Lambda$ takes the
form
 \eqn\gaugemed{W=S f \tilde f,}
which is the standard form of gauge-mediated theories of
supersymmetry breaking, with
 \eqn\vevS{\langle S\rangle \equiv s +
 \theta^2 F_s = m_f + \theta^2 {\Lambda \over M_p} F_{\Phi}
 \simeq {\tilde \Lambda^3 \over M_p^2}+
 \theta^2{\Lambda^2 \tilde \Lambda^3 \over M_p^3}. }
It is easy to check that all other possible terms which we did not
write down explicitly in \superpot\ merely give small corrections
to \vevS\ and do not destabilize the SUSY-breaking vacuum (though
they could shift it by a small amount).

Note that in addition to the SUSY breaking vacuum analyzed above,
this model also has several SUSY-preserving vacua, as expected from
a theory with no R-symmetry \NelsonNF. These vacua include the
SUSY-preserving vacua of the SQCD theory, and they are all far away
from the SUSY-breaking vacuum for the parameters we choose, such
that the lifetime of this vacuum is very long.

As discussed, for instance, in \MurayamaYF, in order for the
analysis above to be valid and in order to find a realistic spectrum
several inequalities need to be satisfied. For the longevity of the
metastable vacuum in the SQCD sector we need
 \eqn\metas{\epsilon = {m_Q \over \Lambda} \simeq {\tilde \Lambda^3
 \over \Lambda M_p^2} \ll 1. }
In order for the messenger fields $f$ not to be tachyonic we require
$F_s\ \roughly< \ |m_f|^2$, or
 \eqn\fnott{ \Lambda^2  M_p\ \roughly<\ \tilde \Lambda^3.}
Finally, in order to preserve the successes of gauge mediation,
including the lack of flavor changing neutral currents, we would
like the gauge-mediated contribution to the scalar masses to be much
larger than other contributions which arise from terms of the form
${1\over M_p^2} Q^\dagger Q v^\dagger v$ in the K\"ahler potential
($v$ is a typical MSSM chiral superfield). This leads to a bound
like
 \eqn\massinemm{ m_{SUSY} \sim {\alpha_{MSSM} \over 4\pi m_f}
 F_s\  \roughly>\  100 \cdot {1\over M_p}F_{\Phi} }
and hence
 \eqn\gaugemed{{\tilde \Lambda^3 \over \Lambda M_p^2}\ \roughly<
 \ 10^{-4},}
 which is a stronger version of \metas.

Expressed in terms of the typical scale expected for SUSY breaking
and the Planck scale
 \eqn\scales{\mu \sim {F_s\over m_f} \sim {\Lambda^2 \over M_p}
  \sim 10^5 GeV  \qquad; \qquad M_p \sim 10^{19} GeV }
we find
 \eqn\finaline{\eqalign{ &\Lambda = \sqrt{\mu M_p} \sim 10^{12}
 GeV,\cr
 &\Lambda^2 M_p \sim  10^{43}GeV^3\  \roughly< \ \tilde \Lambda^3\
 \roughly< \
 10^{-4} \Lambda M_p^2 \sim 10^{46} GeV^3,
 }}
so we can take, for example, $\tilde \Lambda \sim 10^{15} GeV$,
and then $\sqrt{F_s} \sim 10^6 GeV$, $m_Q \sim m_f \sim 10^7 GeV$
and $\epsilon \sim 10^{-5}$.

Although the range in \finaline\ might appear very narrow, in fact
the situation is better. First, by including coupling constants
and various factors of order one in \superpot\ the range in
\finaline\ can be expanded. Second, note that there are no
phenomenological problems if the inequalities \finaline\ are close
to being saturated (hence we used the symbol $\roughly<$).  If the
first one is close to being saturated, the messenger fields could
be lighter than $m_f$, but there is nothing wrong with that, while
the second inequality already includes a safety factor of $100$.

\bigskip
\centerline{\bf Acknowledgements} We would like to thank M.~Dine for
useful comments.  This work was supported in part by grant
\#DE-FG02-90ER40542. The work of O.A. was supported in part by the
Israel-U.S. Binational Science Foundation, by the Israel Science
Foundation, by the European network MRTN-CT-2004-512194, by a grant
from the G.I.F., the German-Israeli Foundation for Scientific
Research and Development, by Minerva, and by a grant of DIP (H.52).

\listrefs
\end